\documentclass[12pt,onecolumn,final]{IEEEtran}
\newlength \figwidth
\usepackage{cite}
\ifCLASSINFOpdf
  \usepackage{graphicx,epstopdf}
  \graphicspath{{../Figures/}}
  \DeclareGraphicsExtensions{.eps}
\else
  \usepackage[dvips]{graphicx}
\fi
\usepackage[cmex10]{amsmath}
\usepackage[tight,footnotesize]{subfigure}
\usepackage{amssymb}
\usepackage{amsthm}
\usepackage{url}
\usepackage{dsfont}

\begin{document}
\newcommand{\indic}{\mathds{1}}
\newcommand{\Ak}{\mathcal{A}_k}
\newcommand{\Bk}{\mathcal{B}_k}
\newcommand{\Dk}{\mathcal{D}_k}
\newcommand{\Ok}{\mathcal{O}_k}
\newcommand{\Pk}{\mathcal{P}_k}
\newcommand{\Ao}{\mathcal{A}^{\circ}}
\newcommand{\Bo}{\mathcal{B}^{\circ}}
\newcommand{\Do}{\mathcal{D}^{\circ}}
\newcommand{\Oo}{\mathcal{O}^{\circ}}
\newcommand{\Po}{\mathcal{P}^{\circ}}
\newcommand{\Ro}{R^{\circ}}
\newcommand{\RBCCk}{R_{\textrm{BCC},k}}
\newcommand{\RBCCo}{R_{\textrm{BCC}}^{\circ}}
\newcommand{\xiBCo}{\xi_{\textrm{BC}}^{\circ}}
\newcommand{\xiBCCo}{\xi_{\textrm{BCC}}^{\circ}}
\newcommand{\xiBCCEo}{\xi_{\textrm{BCCE}}^{\circ}}
\newcommand{\xiBCCEb}{\bar{\xi}_{\textrm{BCCE}}}
\newcommand{\xiBCCEs}{\xi_{\textrm{BCCE}}^{\star}}
\newcommand{\EPhi}{\mathbb{E}_{\Phi_e}}
\newcommand{\gLk}{\gamma_{k}}
\newcommand{\gMk}{\gamma_{M,k}}
\newcommand{\gEk}{\gamma_{E,k}}
\newcommand{\gek}{\gamma_{e,k}}
\newcommand{\gxk}{\gamma_{x,k}}
\newcommand{\gLo}{\gamma^{\circ}}
\newcommand{\gMo}{\gamma_M^{\circ}}
\newtheorem{Theorem}{\bf Theorem}
\newtheorem{Corollary}{\bf Corollary}
\newtheorem{Remark}{\bf Remark}
\newtheorem{Lemma}{\bf Lemma}
\newtheorem{Proposition}{\bf Proposition}
\newtheorem{definition}{\bf Definition}
\title{Secrecy Rates in the Broadcast Channel with Confidential Messages and External Eavesdroppers}
\author{\normalsize\authorblockN{{Giovanni~Geraci$^{1,2}$, Sarabjot~Singh$^{3}$, Jeffrey~G.~Andrews$^{3}$, Jinhong~Yuan$^{1}$, and Iain~B.~Collings$^{2}$}}\\
\small\authorblockA{$^1$School of Electrical Engineering \& Telecommunications,
The University of New South Wales, \textsc{Australia} }\\
\authorblockA{$^2$Wireless and Networking Technologies Laboratory, CSIRO ICT Centre, Sydney, \textsc{Australia}}\\
\authorblockA{$^3$Dept. of Electrical and Computer Engineering, The University of Texas at Austin, \textsc{USA}}
}
\maketitle
%\vspace*{-0.8cm}
\thispagestyle{empty}
\begin{abstract}
In this paper, we consider the broadcast channel with confidential messages and external eavesdroppers (BCCE), where a multi-antenna base station simultaneously communicates to multiple potentially malicious users, in the presence of randomly located external eavesdroppers. Using the proposed model, we study the secrecy rates achievable by regularized channel inversion (RCI) precoding by performing a large-system analysis that combines tools from stochastic geometry and random matrix theory. We obtain explicit expressions for the probability of secrecy outage and an upper bound on the rate loss due to the presence of external eavesdroppers. We show that both these quantities scale as $\frac{\lambda_e}{\sqrt{N}}$, where $N$ is the number of transmit antennas and $\lambda_e$ is the density of external eavesdroppers, irrespective of their collusion strategy. Furthermore, we derive a practical rule for the choice of the regularization parameter, which is agnostic of channel state information and location of eavesdroppers, and yet provides close to optimal performance.
\end{abstract}

\begin{IEEEkeywords}
Physical layer security, broadcast channel, linear precoding, stochastic geometry, random matrix theory.
\end{IEEEkeywords}
\IEEEpeerreviewmaketitle
\newpage
\section{Introduction}

Multiuser multiple-input multiple-output (MIMO) wireless techniques have received tremendous attention as a way to achieve high spectral efficiency in current mobile communication systems such as Long Term Evolution (LTE) \cite{LimMagazine13}. Due to the broadcast nature of the physical medium, wireless multiuser communications are very susceptible to eavesdropping, and it is critical to secure the transmitted information. Security has traditionally been achieved at the network layer with cryptographic
schemes. However, classical cryptography might not be suitable in large dynamic networks, since it requires key distribution and management, and complex encryption/decryption algorithms \cite{Mukherjee10Survey,Shiu11}. A method that exploits the characteristics of wireless channels, such as fading and noise, was proposed as an alternative to achieve perfect secrecy without requiring encryption keys \cite{Wyner75,Csiszar78,LiangBook,LiuBook}. This technique is known as physical layer security, and it has recently become a very active area of research.

\subsection{Motivation and Related Work}

The underlying channel for multiuser MIMO wireless communications is referred to as the MIMO broadcast channel (BC), where a central base station (BS) with $N$ antennas simultaneously communicates to $K$ users over the same frequency band. While it is known that dirty-paper coding (DPC) is a capacity achieving precoding strategy for the Gaussian MIMO BC \cite{CaireZFDPC03}, the non-linearity of the DPC precoder makes it too complex to be implemented \cite{Spencer04Magazine,Li10a}. Linear strategies like regularized channel inversion (RCI) precoding were proposed as a low-complexity alternative for practical systems \cite{Joham05,Peel05,Ryan09}, and their performance was studied by a large-system approach that employs random matrix theory (RMT) tools \cite{NguyenGCOM08,Wagner12}.

Physical layer security was considered to protect the confidentiality of data in the BC, by introducing the broadcast channel with confidential messages (BCC), where the users can act maliciously as eavesdroppers \cite{Ekrem12,Liu13,FakoorianJSAC,Yang11}. A large-system analysis of the secrecy rates achievable by RCI precoding in the BCC was performed by using RMT tools in \cite{Geraci12,GeraciJSAC,GeraciLetter}, where eavesdropping was assumed from the malicious users only. The presence of external eavesdroppers and its effect on the secure connectivity in random wireless networks were studied, among others, in \cite{Haenggi08isit,ZhouGanti11,Pinto12I,Pinto12II} by employing stochastic geometry (SG) tools, but the system model did not account for the potentially malicious behavior of the users.

In a practical scenario, both malicious users and external nodes can act as eavesdroppers. A physical layer security system designed by considering either one of them should be regarded as vulnerable. In fact, a system designed by only considering the presence of external eavesdroppers would be vulnerable to the potential malicious behavior of the users. On the other hand, considering the malicious users only would make the system vulnerable to secrecy outage caused by eavesdropping nodes external to the network. For these reasons, it is of critical importance to study broadcast channels with confidential messages and external eavesdroppers.

\subsection{Approach and Contributions}

In this paper, we introduce the broadcast channel with confidential messages and external eavesdroppers (BCCE) to model a scenario where both (i) malicious users, and (ii) randomly located external nodes can act as eavesdroppers. This is a practical scenario that has not yet been addressed. We study the performance of RCI precoding in the BCCE by performing a large-system analysis that uses results from both SG and RMT. Stochastic geometry is a powerful tool to study a large network with a random distribution of external eavesdroppers \cite{Stoyan96}, whereas random matrix theory enables a deterministic abstraction of the physical layer, for a fixed network topology \cite{CouilletBook}. By combining SG and RMT, we can provide explicit expressions for the average large-system performance with respect to the spatial distribution of the nodes and to the fluctuations of their channels. Our main contributions are summarized below.
\begin{itemize}
\item We obtain the large-system probability of secrecy outage for the RCI precoder in the BCCE, for the two cases of non-colluding and colluding eavesdroppers. We find that the large-system probability of secrecy outage scales as $\frac{\lambda_e}{\sqrt{N}}$, where $N$ is the number of transmit antennas and $\lambda_e$ is the density of external eavesdroppers, irrespective of their collusion strategy.
\item We derive the large-system mean secrecy rate achievable by the RCI precoder in the BCCE. By comparing the mean secrecy rate to the secrecy rate achievable in the BCC, we obtain an upper bound on the rate loss due to the presence of external eavesdroppers, which also scales as $\frac{\lambda_e}{\sqrt{N}}$.
\item We propose a rule for the choice of the regularization parameter $\xi$ of the precoder that maximizes the mean large-system secrecy rate. The function of $\xi$ is to achieve a tradeoff between the signal power at the legitimate user and the crosstalk at the malicious users. The proposed choice of $\xi$ is practical, since it does not require knowledge of either the fluctuations of the channels or the spatial locations of the eavesdroppers, and it provides close to optimal performance.
\end{itemize}

The remainder of the paper is organized as follows. Section II introduces the broadcast channel with confidential messages and external eavesdroppers (BBCE) and the secrecy rates achievable by RCI precoding. In Section III, we derive the probability of secrecy outage, for both cases of non-colluding and colluding external eavesdroppers. In Section IV, we derive the mean secrecy rates achievable by RCI precoding in the BCCE, we study the rate loss due to the presence of external eavesdroppers, and we propose a practical rule for the choice of the regularization parameter of the precoder. In Section V, we provide several numerical results that confirm the accuracy of the analysis. The paper is concluded in Section VI and future work is suggested.
\section{System Model}

In this section, we first recall some results on the MISO BCC, where malicious users connected to the network can act as eavedroppers. Then we introduce the MISO BCCE, where not only malicious users but also nodes external to the network can act as eavesdroppers. This is the case in a real system, where external nodes are randomly scattered in space. These nodes must be regarded as potential eavesdroppers, otherwise the system would be vulnerable to secrecy outage. The BCCE therefore represents a practical scenario that needs to be addressed.

\subsection{Preliminaries: Broadcast Channel with Confidential Messages (BCC)}
We first consider the downlink of a narrowband MISO BCC, consisting of a base station with $N$ antennas which simultaneously transmits $K$ independent confidential messages to $K$ spatially dispersed single-antenna users. In this model, transmission takes place over a block fading channel, and the transmitted signal is $\mathbf{x} = \left[x_1,\ldots,x_N \right]^{T} \in \mathbb{C}^{N \times 1}$. We assume homogeneous users, i.e., each user experiences the same received signal power on average, thus the model assumes that their distances from the transmitter are the same and unitary. The received signal at user $k$ is given by
\begin{equation}
y_k=\sum_{j=1}^{N} h_{k,j}x_{j}+n_{k}
\label{eqn:MIMO_scalar}
\end{equation}
where $h_{k,j} \sim \mathcal{CN}(0,1)$ is the i.i.d. channel between the $j^{\textrm{th}}$ transmit antenna element and the $k^{\textrm{th}}$ user, and $n_{k} \sim \mathcal{CN}(0,\sigma^2)$ is the noise seen at the $k^{\textrm{th}}$ receiver. 
The corresponding vector equation is
\begin{equation}
\mathbf{y}=\mathbf{Hx}+\mathbf{n}
\label{eqn:MIMO_vector}
\end{equation}
where $\mathbf{H} = \left[\mathbf{h}_1,\ldots,\mathbf{h}_K \right]^{\dagger}$ is the $K \times N$ channel matrix. We assume $\mathbb{E}[ \mathbf{nn}^{\dagger} ] =\sigma^{2} \mathbf{I}_K$, where $\mathbf{I}_K$ is the $K \times K$ identity matrix, define the SNR $\rho \triangleq 1/ \sigma ^2$, and impose the long-term power constraint $\mathbb{E}[ \left\| \mathbf{x} \right\|^{2} ] =1$. For each user $k$, we denote by $\mathcal{M}_k = \left\{ 1,\ldots,k-1,k+1,\ldots,K\right\}$ the set of remaining users. In general, the behavior of the users cannot be determined by the BS. As a worst-case scenario, we assume that for each user $k$, all users in $\mathcal{M}_k$ can cooperate to jointly eavesdrop on the $k^{\textrm{th}}$ message. Since the set of malicious users $\mathcal{M}_k$ can perform joint processing, they can be seen as a single equivalent malicious user $M_k$ with $K-1$ receive antennas.

In this paper, we consider regularized channel inversion (RCI) precoding. In RCI precoding, the transmitted vector $\mathbf{x}$ is obtained at the BS by performing a linear processing on the vector of confidential messages $\mathbf{u} = \left[u_1,\ldots,u_K \right]^{T}$, whose entries are chosen independently, satisfying $\mathbb{E}[ \left|u_k\right|^2 ] =1$. The transmitted signal $\mathbf{x}$ after RCI precoding can be written as $\mathbf{x} = \mathbf{Wu}$, where $\mathbf{W} = \left[\mathbf{w}_1,\ldots,\mathbf{w}_K \right]$ is the $N \times K$ RCI precoding matrix, given by \cite{Peel05,NguyenGCOM08,Wagner12}
\begin{equation}
\mathbf{W} = \frac{1}{\sqrt{\zeta}} \mathbf{H}^{\dagger} \left( \mathbf{H H}^{\dagger} + N \xi \mathbf{I}_K \right) ^{-1} = \frac{1}{\sqrt{\zeta}} \left( \mathbf{H}^{\dagger} \mathbf{H} + N \xi \mathbf{I}_N \right) ^{-1} \mathbf{H}^{\dagger}
\label{eqn:RCI_precoder}
\end{equation}
and $\zeta = \textrm{tr} \left\{ \mathbf{H}^{\dagger} \mathbf{H} \left( \mathbf{H}^{\dagger} \mathbf{H} + N \xi \mathbf{I}_N \right) ^{-2} \right\}$ is a long-term power normalization constant. The function of the real regularization parameter $\xi$ is to achieve a tradeoff between the signal power at the legitimate user and the interference and information leakage at the other unintended users for each message.

Due to cooperation, interference cancellation can be performed at the equivalent malicious user $M_k$, which does not see any undesired signal term apart from the received noise. As a result, a secrecy rate achievable for user $k$ by RCI precoding is given by \cite{Geraci12}
\begin{equation}
\RBCCk = \left[ \log_2 \Big( 1 + \gLk \Big) - \log_2 \Big( 1 + \gMk \Big) \right]^+,
\label{eqn:Rs}
\end{equation}
where we use the notation $[ \cdot ]^+ \triangleq \max(\cdot,0)$, and where $\gLk$ and $\gMk$ are the signal-to-interference-plus-noise ratios for the message $u_k$ at the legitimate receiver $k$ and the equivalent malicious user $M_k$, respectively, given by
\begin{equation}
\gLk = \frac {\rho \left| \mathbf{h}_k^{\dagger} \mathbf{w}_k \right| ^2} {1 + \rho \sum_{j \neq k} {\left| \mathbf{h}_k^{\dagger} \mathbf{w}_j \right| ^2} } \quad \textrm{and} \quad
\gMk = \rho \left\| \mathbf{H}_k \mathbf{w}_{k} \right\| ^2,
\label{eqn:SINR}
\end{equation}
and where $\mathbf{H}_k$ is the matrix obtained from $\mathbf{H}$ by removing the $k^{\textrm{th}}$ row.

The secrecy rate of the RCI precoder in the large-system regime was studied in \cite{GeraciJSAC}, where both the number of receivers $K$ and the number of transmit antennas $N$ approach infinity, with their ratio $\beta = K/N$ being held constant. The value of $\beta$ represents the network load. Let $\rho>0$, $\beta>0$, and let $\RBCCk$ be the secrecy rate achievable by RCI precoding in the BCC defined in (\ref{eqn:Rs}). Then \cite{GeraciJSAC}
\begin{equation}
\left| \RBCCk - R_{\textrm{BCC}}^{\circ} \right| \stackrel{\textrm{a.s.}}{\longrightarrow} 0, \quad \textrm{as} \quad N \rightarrow \infty, \quad \forall k
\label{eqn:Theorem1}
\end{equation}
where $R_{\textrm{BCC}}^{\circ}$ denotes the secrecy rate in the large-system regime, given by
\begin{equation}
R_{\textrm{BCC}}^{\circ} = \left[ \log_2 \frac{1+\gLo}
{1+\gMo} \right]^+,
\label{eqn:Rs_large_system}
\end{equation}
and where 
\begin{equation}
\gLo = g\left( \beta,\xi \right)
\frac{\rho + \frac{\rho\xi}{\beta} \left[ 1 + g\left( \beta,\xi \right) \right] ^2 
}{\rho + \left[ 1 + g\left( \beta,\xi \right) \right] ^2}	\quad \textrm{and} \quad \gMo = \frac{\rho}{ \left[ 1+g\left( \beta,\xi \right) \right] ^2},
\label{eqn:gamma_circ}
\end{equation}
with 
$g \left( \beta,\xi \right) = \frac{1}{2} \left[ \sqrt{ \frac{\left(1-\beta \right)^2}{\xi^2}  +  \frac{2\left(1+\beta\right)}{\xi}  +  1} +  \frac{1-\beta}{\xi}  -  1 \right]$. The optimal value of $\xi$ that maximizes the large-system secrecy rate $R_{\textrm{BCC}}^{\circ}$ was obtained in \cite{GeraciJSAC} and it is given by
\begin{equation}
\xiBCCo = \frac {-2\rho^2\left( 1 - \beta \right)^2 + 6\rho \beta + 2\beta^2 - 2 \left[ \beta \left( \rho+1 \right) -\rho \right] \cdot \sqrt{ \beta^2 \left[ \rho^2 + \rho + 1 \right] - \beta \left[ 2 \rho \left( \rho -1 \right) \right] + \rho^2 } } {6 \rho^2 \left( \beta + 2 \right) + 6 \rho \beta}.
\label{eqn:xi_opt}
\end{equation}

\subsection{Broadcast Channel with Confidential Messages and External Eavesdroppers (BCCE)}

\begin{figure}
\centering
\includegraphics[width=\figwidth]{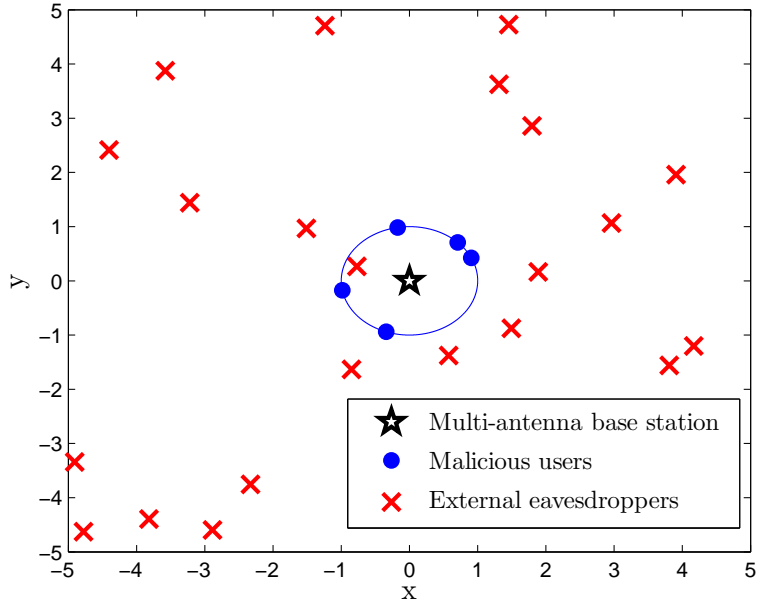}
\caption{Example of a BCCE with $K=5$ malicious users and a density of external eavesdroppers $\lambda_e=0.2$.}
\label{fig:PPP}
\end{figure}

We now consider the MISO BCCE, by including external single-antenna eavesdroppers in the system. The external eavesdroppers are assumed to be distributed on the two-dimensional plane according to a Poisson point process (PPP) $\Phi_e$ of density $\lambda_e$ \cite{Stoyan96}. Fig. \ref{fig:PPP} shows an example of BCCE, where the BS is at the origin, and the users lie on a disc of radius $1$. As a worst-case scenario, we assume that each eavesdropper can cancel the interference caused by the remaining $K-1$ messages. Assuming that the BS lies at the origin, the SINR $\gek$ for the $k^{\textrm{th}}$ message at a generic eavesdropper located in $e$ is then given by
\begin{equation}
\gek = \frac{\left| \mathbf{h}_e^{\dagger} \mathbf{w}_k \right|^2}{\|e\|^{\eta}\sigma^2}
\end{equation}
where $\mathbf{h}_e^{\dagger}$ is the channel vector between the base station and the eavesdropper in $e$, and it takes into account the Rayleigh fading, and $\eta$ is the path loss exponent. Some of the results provided in this paper assume a path loss exponent $\eta=4$. In this special case, which is a reasonable value for $\eta$ in a shadowed urban area \cite{Rappaport}, it is possible to obtain compact expressions for quantities of interest, such as the probability of secrecy outage and the mean secrecy rate.

The precoding vector $\mathbf{w}_k$ is calculated independently of $\mathbf{h}_e^{\dagger}$, therefore they are independent isotropic random vectors. The channel $\mathbf{h}_e^{\dagger}$ has unit norm, whereas the precoding vector $\mathbf{w}_k$ has norm $\frac{1}{\sqrt{K}}$ because it is obtained after the normalization $\|\mathbf{W}\|^2 = \sum_{k=1}^{K} {\|\mathbf{w}_k\|^2} = 1$. The inner product $\mathbf{h}_e^{\dagger} \mathbf{w}_k$ is a linear combination of $N$ complex normal random variables, therefore $\left| \mathbf{h}_e^{\dagger} \mathbf{w}_k \right|^2 \sim \textrm{exp}(\frac{1}{K})$.

In the following, we consider two types of external eavesdroppers, namely non-colluding eavesdroppers and colluding eavesdroppers. In the non-colluding case, the eavesdroppers individually overhear the communication without centralized processing. In the colluding eavesdroppers case, all eavesdroppers are able to jointly process their received message at a central data processing unit. The secrecy rate $R_{k}$ achievable by the $k^{\textrm{th}}$ user in the BCCE is given by
\begin{equation}
R_{k} = \left[ \log_2 \Big( 1 + \gLk \Big) - \log_2 \Big( 1 + \max \left(\gMk,\gEk\right) \Big) \right]^+,
\label{eqn:Rs_e}
\end{equation}
where $\gEk$ is the resulting SINR of the PPP of external eavesdoppers for the $k^{\textrm{th}}$ message. The secrecy rate $R_k$ is therefore affected by the maximum of the SINR $\gMk$ at the alliance of malicious users and the SINR $\gEk$ at the external eavesdroppers. In the case of non-colluding eavesdroppers, $\gEk$ is the SINR at the strongest eavesdropper. In the case of colluding eavesdroppers, all eavesdroppers can perform joint processing, and they can, therefore, be seen as a single multi-antenna eavesdropper. After interference cancellation, each eavesdropper receives the useful signal embedded in noise, and the optimal receive strategy at the colluding eavesdroppers is maximal ratio combining (MRC) which yields to an SINR $\gEk = \sum_{e \in \Phi_e} \gek$ given by the sum of the SINRs $\gek$ at all eavesdroppers.

The achievable secrecy sum-rate is denoted by $S$ and defined as $S = \sum_{k=1}^{K} R_k$.
\section{Probability of Secrecy Outage}

In this section, we derive the secrecy outage probability, i.e., the probability that the secrecy rate $R_k$ achievable by user $k$ with RCI precoding in the BCCE is zero, for both cases of non-colluding and colluding eavesdroppers. Then we study the secrecy outage probability in the large-system regime, and determine how the number of antennas $N$ must scale in order to guarantee a given secrecy outage probability. The secrecy outage probability for user $k$ is defined as
\begin{equation}
  \Ok \triangleq \mathbb{P}(R_k = 0) = \left\{ 
  \begin{array}{c c l}
  1 & \quad \textrm{if} \enspace \gLk \leq \gMk\\
  \mathbb{P}(\gEk \geq \gLk \, | \, \gLk)  & \quad \textrm{otherwise}\\
  \end{array} \right.
\label{eqn:outage_definition}
\end{equation}
In most cases, RCI precoding ensures $\gLk > \gMk$ \cite{GeraciJSAC}, and therefore, the secrecy outage probability is often given by the probability that $R_k$ is driven to zero by the presence of external eavesdroppers.

\subsection{Non-colluding Eavesdroppers}

In the case of non-colluding eavesdroppers, $\gEk$ is the SINR at the strongest eavesdropper $E$, given by
\begin{equation}
\gEk = \max_{e \in \Phi_e} \!\enspace \gek = \max_{e \in \Phi_e} \!\enspace \frac{\left| \mathbf{h}_e^{\dagger} \mathbf{w}_k \right|^2}{\|e\|^{\eta}\sigma^2}.
\end{equation}

In the case of non-colluding eavesdroppers, $\Ok$ is the probability that any eavesdropper has an SINR greater than or equal to the SINR of the legitimate user $k$. We obtain the following result.
\begin{Lemma}
The secrecy outage probability for user $k$ in the presence of non-colluding eavesdroppers is given by
\begin{equation}
  \Ok = \left\{ 
  \begin{array}{c c l}
  1 & \quad \textrm{if} \enspace \gLk \leq \gMk\\
  1 - \mathrm{exp}\left[ - \frac{2 \pi \lambda_e \Gamma\left(\frac{2}{\eta}\right) }{\eta (N \beta \sigma^2 \gLk)^{\frac{2}{\eta}}}\right]  & \quad \textrm{otherwise}\\
  \end{array} \right.
\label{eqn:Pn}
\end{equation}
\label{Lemma1}
where $\Gamma(\cdot)$ is the gamma function defined as
\begin{equation}
\Gamma(z) \triangleq \int_0^{\infty} {t^{z-1} e^{-t} dt}.
\end{equation}
\end{Lemma}
\begin{IEEEproof}
See Appendix A.
\end{IEEEproof}
By applying results from RMT \cite{CouilletBook}, we now obtain the large-system secrecy outage probability $\Oo$ in the presence of non-colluding eavesdroppers.
\begin{Theorem}
The secrecy outage probability in the presence of non-colluding eavesdroppers satisfies
\begin{equation}
| \Ok - \Oo | \stackrel{a.s.}{\longrightarrow} 0 , \quad \textrm{as} \enspace N \rightarrow \infty, \quad \forall k
\end{equation}
where
\begin{equation}
  \Oo = \left\{ 
  \begin{array}{c c l}
  1 & \quad \textrm{if} \enspace \gLo \leq \gMo\\
  1 - \mathrm{exp}\left[ - \frac{2 \pi \lambda_e \Gamma\left(\frac{2}{\eta}\right) }{\eta (N \beta \sigma^2 \gLo)^{\frac{2}{\eta}}}\right]  & \quad \textrm{otherwise}\\
  \end{array} \right.
\label{eqn:Pn_large}
\end{equation}
\end{Theorem}
\begin{IEEEproof}
Theorem 1 follows from Lemma \ref{Lemma1}, by noting that $|\gLk - \gLo| \stackrel{a.s.}{\longrightarrow} 0$ as $N \rightarrow \infty$, and by the continuous mapping theorem \cite{BillingsleyBook}.
\end{IEEEproof}

\begin{Corollary}
If $\gLo > \gMo$ and $\eta=4$, then (i) the number of transmit antennas required in order to guarantee a large-system secrecy outage probability $\Oo<\epsilon$ in the presence of non-colluding eavesdroppers is $N > \left(\frac{\mu \lambda_e}{\epsilon \sqrt{\gLo}}\right)^2$, where $\mu \triangleq \frac{\pi^{\frac{3}{2}}}{2 \sqrt{\beta \sigma^2}}$, and (ii) the large-system secrecy outage probability $\Oo$ decays as $\frac{1}{\sqrt{N}}$.
\end{Corollary}
\begin{IEEEproof}
The proof follows from Theorem 1, by noting that $\Gamma\left(\frac{1}{2}\right)=\sqrt{\pi}$, and that $1-e^{-x} > x$ for $0 < x < 1$.
\end{IEEEproof}

A special case of the previous scenario is the one where only the eavesdropper which is nearest to the base station attempts to eavesdrop. In this case we have
\begin{equation}
\gEk = \frac{\left| \mathbf{h}_E^{\dagger} \mathbf{w}_k \right|^2}{\|E\|^{\eta}\sigma^2}
\end{equation}
where
\begin{equation}
E = \underset{e \in \Phi_e}{\textrm{argmin}} \, \|e\|.
\end{equation}
\begin{Lemma}
The secrecy outage probability for user $k$, caused by the external eavesdropper nearest to the base station, under a path loss exponent $\eta=4$, is given by
\begin{equation}
  \Ok = \left\{ 
  \begin{array}{c c l}
  1 & \quad \textrm{if} \enspace \gLk \leq \gMk\\
  \frac{2 \mu \lambda_e}{\sqrt{N \gLk}} \mathrm{exp} \left( \frac{ \mu^2 \lambda_e^2}{\pi N \gLk } \right) \mathrm{Q} \left( \mu \lambda_e \sqrt{\frac{2}{\pi N \gLk}} \right)  & \quad \textrm{otherwise}\\
  \end{array} \right.
\label{eqn:P1}
\end{equation}
\label{Lemma2}
where $\mathrm{Q}(\cdot)$ is the Q-function defined as
\begin{equation}
\mathrm{Q}(x) \triangleq \frac{1}{\sqrt{2 \pi}} \int_x^{\infty} \mathrm{exp}\left(-\frac{u^2}{2}\right)\, du.
\end{equation}
\end{Lemma}
\begin{IEEEproof}
See Appendix B.
\end{IEEEproof}
By applying results from RMT, we now obtain the large-system secrecy outage probability $\Oo$ caused by the eavesdropper which is nearest to the base station.
\begin{Theorem}
The secrecy outage probability for user $k$, caused by the external eavesdropper nearest to the base station, under a path loss exponent $\eta=4$, satisfies
\begin{equation}
| \Ok - \Oo | \stackrel{a.s.}{\longrightarrow} 0 , \quad \textrm{as} \enspace N \rightarrow \infty, \quad \forall k
\end{equation}
where
\begin{equation}
  \Oo = \left\{ 
  \begin{array}{c c l}
  1 & \quad \textrm{if} \enspace \gLo \leq \gMo\\
  \frac{\mu \lambda_e}{\sqrt{N}} \left( 1 + \frac{\mu^2 \lambda_e^2}{\pi N} \right) \left( 1 - \frac{2 \mu \lambda_e}{\pi \sqrt{N}} \right)  & \quad \textrm{otherwise}\\
  \end{array} \right.
\label{eqn:P1_large}
\end{equation}
\end{Theorem}
\begin{IEEEproof}
Theorem 2 follows from Lemma \ref{Lemma2}, by first-order Taylor approximation of (\ref{eqn:P1}), by noting that $|\gLk - \gLo| \stackrel{a.s.}{\longrightarrow} 0$ as $N \rightarrow \infty$, and by the continuous mapping theorem \cite{BillingsleyBook}.
\end{IEEEproof}

\subsection{Colluding Eavesdroppers}

The colluding eavesdroppers case represents a worst-case scenario. In this case, all eavesdroppers can perform joint processing, and they can therefore be seen as a single multi-antenna eavesdropper. After interference cancellation, each eavesdropper receives the useful signal embedded in noise, and the optimal receive strategy at the colluding eavesdroppers is maximal ratio combining (MRC). This yields to an SINR $\gEk$ at the colluding eavesdroppers given by
\begin{equation}
\gEk = \frac{1}{\sigma^2} \sum_{e \in \Phi_e} \|e\|^{-\eta} \left| \mathbf{h}_e^{\dagger} \mathbf{w}_k \right|^2.
\end{equation}

\begin{Lemma}
The secrecy outage probability for user $k$ in the presence of colluding eavesdroppers, under a path loss exponent $\eta=4$, is given by
\begin{equation}
  \Ok = \left\{ 
  \begin{array}{c c l}
  1 & \quad \textrm{if} \enspace \gLk \leq \gMk\\
  1 - 2\mathrm{Q} \left( \mu \lambda_e \sqrt{\frac{\pi}{2 N \gLk}} \right)   & \quad \textrm{otherwise}\\
  \end{array} \right.
\label{eqn:Pc}
\end{equation}
\label{Lemma3}
\end{Lemma}
\begin{IEEEproof}
See Appendix C.
\end{IEEEproof}
By applying results from RMT, we now obtain the large-system secrecy outage probability $\Oo$ in the presence of colluding eavesdroppers.
\begin{Theorem}
The secrecy outage probability in the presence of colluding eavesdroppers, under a path loss exponent $\eta=4$, satisfies
\begin{equation}
| \Ok - \Oo | \stackrel{a.s.}{\longrightarrow} 0 , \quad \textrm{as} \enspace N \rightarrow \infty, \quad \forall k
\end{equation}
where
\begin{equation}
  \Oo = \left\{ 
  \begin{array}{c c l}
  1 & \quad \textrm{if} \enspace \gLo \leq \gMo\\
  1 - 2\mathrm{Q} \left( \mu \lambda_e \sqrt{\frac{\pi}{2 N \gLo}} \right) 
  & \quad \textrm{otherwise}\\
  \end{array} \right.
\label{eqn:Pc_large}
\end{equation}
\end{Theorem}
\begin{IEEEproof}
Theorem 3 follows from Lemma \ref{Lemma3}, by noting that $\Gamma\left(\frac{1}{2}\right)=\sqrt{\pi}$, that $|\gLk - \gLo| \stackrel{a.s.}{\longrightarrow} 0$ as $N \rightarrow \infty$, and by the continuous mapping theorem \cite{BillingsleyBook}.
\end{IEEEproof}
\begin{Corollary}
Let $\gLo > \gMo$ and $\eta=4$, then (i) the number of transmit antennas required in order to guarantee a large-system secrecy outage probability $\Oo<\epsilon$ in the presence of colluding eavesdroppers is $N > \left(\frac{\mu \lambda_e}{\epsilon \sqrt{\gLo}}\right)^2$, and (ii) the large-system outage probability $\Oo$ decays as $\frac{1}{\sqrt{N}}$.
\end{Corollary}
\begin{IEEEproof}
The proof follows from Theorem 3 and by using $1-2Q(x) < \sqrt{\frac{2}{\pi}}x$ for $0 < x < 1$.
\end{IEEEproof}
\begin{Remark}
By comparing the results in Corollary 1 and Corollary 2, we can conclude that (i) the collusion among eavesdroppers does not significantly affect the number of transmit antennas $N$ required to meet a given probability of secrecy outage in the large-system regime, and (ii) increasing the density of eavesdroppers $\lambda_e$ by a factor $n$ requires increasing $N$ by a factor $n^2$ in order to meet a given probability of secrecy outage.
\end{Remark}
\section{Mean Secrecy Rates}

In this section, we derive the mean secrecy rates, averaged over the location of the external eavesdroppers, achievable by RCI precoding in the BCCE, for both cases of non-colluding and colluding eavesdroppers. We then study the mean secrecy rates in the large-system regime, and derive a bound on the secrecy rate loss due to the presence of external eavesdroppers. Finally, we propose a rule for the choice of the regularization parameter of the precoder that maximizes the mean of the large-system secrecy rate.

\subsection{Mean Secrecy Rate}

We now obtain the following result for the mean secrecy rate at user $k$.
\begin{Lemma}
The mean secrecy rate achievable at user $k$ by RCI precoding in the BCCE is given by
\begin{equation}
  \EPhi \left[ R_k \right] = \left\{ 
  \begin{array}{c c l}
  0 & \quad \textrm{if} \enspace \gLk \leq \gMk \\
  \log_2 \frac{\left(1+\gLk\right)^{1-\Ok}}
{\left(1+\gMk\right)^{1-\Pk}} - \int_{\gMk}^{\gLk} {\log_2(1+y) f_{\gEk}(y) \, dy} & \quad \textrm{otherwise}\\
  \end{array} \right.
\label{eqn:Rk_bar}  
\end{equation}
In (\ref{eqn:Rk_bar}), $\Pk$ is the probability that the SINR $\gamma_{E,k}$ at the external eavesdroppers is greater than or equal to the SINR $\gamma_{M,k}$ at the malicious users, and for a path loss exponent $\eta=4$ is given by
\begin{equation}
  \Pk \triangleq \mathbb{P}(\gEk \geq \gMk) = 
 \left\{ 
  \begin{array}{c c l}
  1 - \mathrm{exp} \left( -\frac{\mu \lambda_e}{\sqrt{N \gMk}} \right)  & \quad \textrm{for non-colluding eavesdroppers} \\
  1-2\mathrm{Q} \left( \mu \lambda_e \sqrt{\frac{\pi}{2 N \gMk}} \right)  & \quad \textrm{for colluding eavesdroppers}\\
  \end{array} \right.
  \label{eqn:Pk}
\end{equation}
and $f_{\gEk}(y)$ is the distribution of the SINR at the external eavesdroppers, given by
\begin{equation}
  f_{\gEk}(y) = \left\{ 
  \begin{array}{c c l}
  \frac{\mu \lambda_e y^{-\frac{3}{2}}}{2 \sqrt{N}} \mathrm{exp} \left( -\frac{\mu  \lambda_e}{\sqrt{N y}} \right) & \quad \textrm{for non-colluding eavesdroppers} \\
  \frac{\mu \lambda_e y^{-\frac{3}{2}}}{2 \sqrt{N}} \mathrm{exp} \left( -\frac{\pi \mu^2 \lambda_e^2}{4 N y} \right) & \quad \textrm{for colluding eavesdroppers}\\
  \end{array} \right.
\end{equation}
\label{Lemma4}
\end{Lemma}
\begin{IEEEproof}
See Appendix D.
\end{IEEEproof}
By applying results from RMT, we now obtain the large-system mean secrecy rate $\Ro$ achievable by RCI precoding in the BCCE.
\begin{Theorem}
The mean secrecy rate achievable for user $k$ by RCI precoding in the BCCE satisfies
\begin{equation}
\left| \EPhi [R_k] - \Ro \right| \stackrel{\textrm{a.s.}}{\longrightarrow} 0, \quad \textrm{as} \quad N \rightarrow \infty, \quad \forall k.
\end{equation}
$\Ro$ denotes the mean secrecy rate in the large-system regime, given by
\begin{equation}
  \Ro = \left\{ 
  \begin{array}{c c l}
  0 & \quad \textrm{if} \enspace \gLo \leq \gMo\\
  \log_2 \frac{\left(1+\gLo\right)^{1-\Oo}}
{\left(1+\gMo\right)^{1-\Po}} - \int_{\gMo}^{\gLo} {\log_2(1+y) f_{\gEk}(y) \, dy} & \quad \textrm{otherwise}\\
  \end{array} \right.
\label{eqn:Rs_BCCE}
\end{equation}
In (\ref{eqn:Rs_BCCE}), $\Po$ is the probability that the SINR $\gamma_{E,k}$ at the external eavesdroppers is greater than or equal to the large-system SINR $\gMo$ at the malicious users, and for $\eta=4$ it is given by
\begin{equation}
  \Po \triangleq \mathbb{P}(\gEk \geq \gMo) = \left\{ 
  \begin{array}{c c l}
  1 - \mathrm{exp} \left( - \frac{\mu \lambda_e}{\sqrt{N \gMo}} \right) & \quad \textrm{for non-colluding eavesdroppers} \\
  1-2\mathrm{Q} \left( \mu \lambda_e \sqrt{\frac{\pi}{2 N \gMo}} \right) & \quad \textrm{for colluding eavesdroppers}\\
  \end{array} \right.
\label{eqn:P}  
\end{equation}
\label{Theorem4}
\end{Theorem}
\begin{IEEEproof}
Theorem 4 follows from Lemma 4, by replacing $\gLk$ and $\gMk$ with their respective deterministic equivalents $\gLo$ and $\gMo$, by applying the continuous mapping theorem, the Markov inequality, and the Borel-Cantelli lemma \cite{BillingsleyBook}.
\end{IEEEproof}

\subsection{Secrecy Rate Loss due to the External Eavesdroppers}

By comparing the large-system mean secrecy rate of the BCCE in (\ref{eqn:Rs_BCCE}) to the large-system secrecy rate of the BCC without external eavesdroppers in (\ref{eqn:Rs_large_system}), for a given regularization parameter $\xi$, we can evaluate the secrecy rate loss $\Delta_e$ due to the presence of external eavesdroppers, defined as
\begin{equation}
\Delta_e \triangleq \RBCCo - \Ro.
\end{equation}
We now obtain an upper bound on the secrecy rate loss $\Delta_e$.
\begin{Corollary}
The secrecy rate loss $\Delta_e$ due to the presence of external eavesdroppers satisfies
\begin{equation}
\Delta_e \leq \Delta_e^{UB} \triangleq \frac{\nu \lambda_e}{\sqrt{N}},
\end{equation}
where $\nu$ is a constant independent of $N$, $\lambda_e$, and of the cooperation strategy at the eavesdroppers, given by
\begin{equation}
\nu = \mu \left[ \frac{\RBCCo}{\sqrt{\gLo}} + \left( \sqrt{\gLo} - \sqrt{\gMo} \right)^+ \right].
\end{equation}
\end{Corollary}
\begin{IEEEproof}
See Appendix E.
\end{IEEEproof}
\begin{Remark}
It follows from Corollary 3 that, irrespective of the collusion strategy at the external eavesdroppers, (i) as the number $N$ of transmit antennas grows, the secrecy rate loss $\Delta_e$ tends to zero as $\frac{1}{\sqrt{N}}$, and (ii) increasing the density of eavesdroppers $\lambda_e$ by a factor $n$ requires increasing $N$ by a factor $n^2$ in order to meet a given value of $\Delta_e^{UB}$.
\end{Remark}

\subsection{Optimal Regularization Parameter}
The value of the regularization parameter $\xi$ has a significant impact on the secrecy rates. The optimal large-system regularization parameter of the RCI precoder for the MISO broadcast channel (BC) without secrecy requirements is given by $\xiBCo = \frac{\beta}{\rho}$ \cite{Peel05,Wagner12,NguyenGCOM08}. The optimal large-system regularization parameter for the MISO broadcast channel with confidential messages (BCC) was derived in \cite{GeraciJSAC} and it is also a function of $\beta$ and $\rho$, given by $\xiBCCo$ in (\ref{eqn:xi_opt}). In the MISO broadcast channel with confidential messages and external eavesdroppers (BCCE), we denote by $\xiBCCEo$ the regularization parameter that maximizes the large-system mean secrecy rate. The value of $\xiBCCEo$ can be obtained by numerically solving the following equation
\begin{equation}
\xiBCCEo \triangleq \underset{\xi}{\textrm{argmax}} \, \Ro
\label{eqn:xi_BCCE}
\end{equation}
with $\Ro$ given in (\ref{eqn:Rs_BCCE}). Since the secrecy rate of the MISO BCCE is affected by the SINR at the external eavesdroppers, the optimal large-system regularization parameter $\xiBCCEo$ is not just a function of $\beta$ and $\rho$, but it also depends on the number of transmit antennas $N$, the density of the eavesdroppers $\lambda_e$, and their collusion strategy. The value of $\xiBCCEo$ should be found as a compromise between: (i) maximizing the SINR $\gLo$ at the legitimate user, and (ii) trading off the SINR $\gMo$ at the malicious users and the probability $\Po$ that the external eavesdroppers are more harmful than the malicious users. We have the following two extreme cases.

\begin{Lemma}
The optimal large-system regularization parameter $\xiBCCEo$ follows the trend:
\begin{equation}
  \begin{array}{c c l}
  &\xiBCCEo \rightarrow \xiBCCo & \quad \textrm{as} \enspace \lambda_e \rightarrow 0\\
  &\xiBCCEo \rightarrow \xiBCo=\frac{\beta}{\rho} & \quad \textrm{as} \enspace \lambda_e \rightarrow \infty\\
  \end{array}
\end{equation}
\end{Lemma}
\begin{IEEEproof}
For low densities $\lambda_e$, we have by Corollary 3 that $\Ro$ approaches $\RBCCo$, therefore $\xiBCCEo$ approaches $\xiBCCo$. For high densities $\lambda_e$, we have $\Pk = \mathbb{P} (\gamma_{E,k} \geq \gamma_{M,k}) \rightarrow 1$, and the secrecy rate $R_k$ in (\ref{eqn:Rs_e}) is determined solely by $\gLk$ and $\gEk$. Since $\gEk$ does not depend on $\xi$, maximizing the mean rate coincides with the rate maximization problem for the BC, and its solution in the large-system regime is given by $\xiBCo$.
\end{IEEEproof}
\section{Numerical Results}

In this section, we provide numerical results to show the performance of RCI precoding in the BCCE, under a path loss exponent $\eta=4$. We consider finite-size systems, and simulate the probability of secrecy outage, the secrecy rate, and the optimal regularization parameter of the precoder, in different scenarios and under different system dimensions, network loads, SNRs, and densities of eavesdroppers. The simulations show that many results obtained in Section III and Section IV by using random matrix theory and stochastic geometry tools hold even for networks with a small number of users and antennas and randomly located eavesdroppers.

\begin{figure}
\centering
\includegraphics[width=\figwidth]{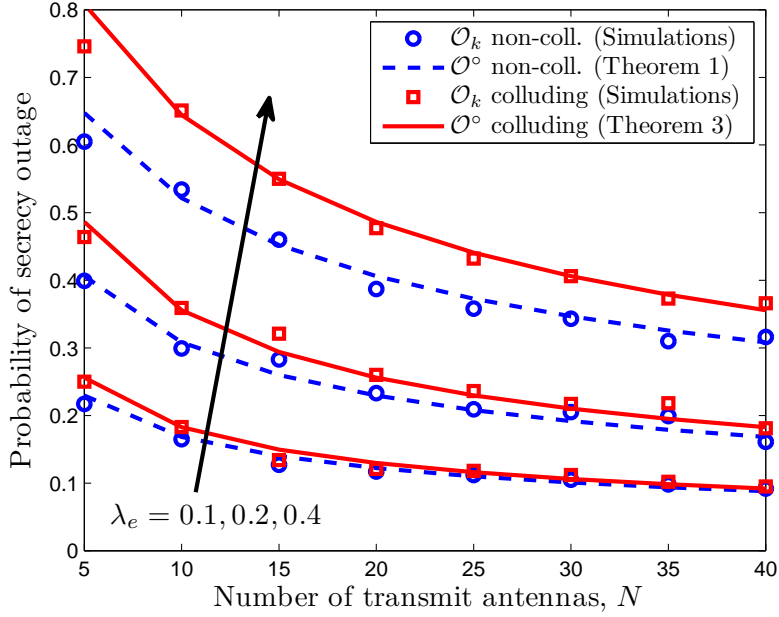}
\caption{Comparison between the simulated probability of outage $\Ok$ and the large-system results $\Oo$ provided in Theorem 1 and Theorem 3, for a network load $\beta=1$, an SNR $\rho=10$dB, and various values of $\lambda_e$.}
\label{fig:SG_RMT}
\end{figure}

In Fig. \ref{fig:SG_RMT} we compare the simulated probability of outage $\Ok$ under non-colluding and colluding eavesdroppers, respectively, to the large-system results $\Oo$ provided in Theorem 1 and Theorem 3, respectively. In the simulations, the regularization parameter $\xiBCCo$ in (\ref{eqn:xi_opt}) was used. We observe that for $\lambda_e=0.1$ and small probabilities of secrecy outage, (i) $N > \left(\frac{\mu \lambda_e}{0.1 \sqrt{\gLo}}\right)^2 = 34$ yields to a secrecy outage probability smaller than $0.1$, (ii) the secrecy outage probability decays as $\frac{1}{\sqrt{N}}$, and (iii) the collusion of eavesdroppers does not significantly affect the probability of secrecy outage. All these observations are consistent with Corollary 1, Corollary 2, and Remark 1.

\begin{figure}
\centering
\includegraphics[width=\figwidth]{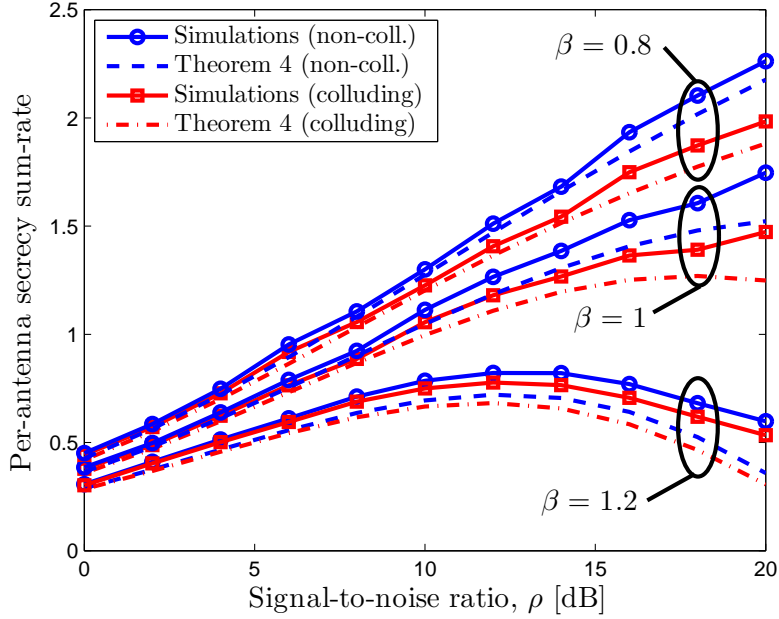}
\caption{Comparison between the simulated ergodic per-antenna secrecy sum-rate $\mathbb{E}[S]/N$ under non-colluding and colluding eavesdroppers, and the large-system results $K \Ro / N$ from Theorem 4, for $\lambda_e=0.1$, $N=10$ transmit antennas, and various values of the network load $\beta$.}
\label{fig:SG_Rs_vs_SNR}
\end{figure}

In Fig. \ref{fig:SG_Rs_vs_SNR} we compare the simulated ergodic per-antenna secrecy sum-rate under non-colluding and colluding eavesdroppers, to the large-system results from Theorem 4, for $\lambda_e=0.1$, $N=10$, $\xi=\xiBCCo$, and various values of $\beta$. We note that the accuracy of the large-system analysis decreases with the SNR. The loss of accuracy is due to the limitations of the tools used from RMT \cite{Wagner12}. Moreover, we note that the per-antenna secrecy sum-rate does not monotonically increase with the SNR. This is due to the fact that in the worst-case scenario the malicious users and the external eavesdroppers can cancel the interference, whereas the legitimate user is interference-limited in the high-SNR regime. This is consistent with the case of BCC \cite{GeraciJSAC}.

\begin{figure}
\centering
\includegraphics[width=\figwidth]{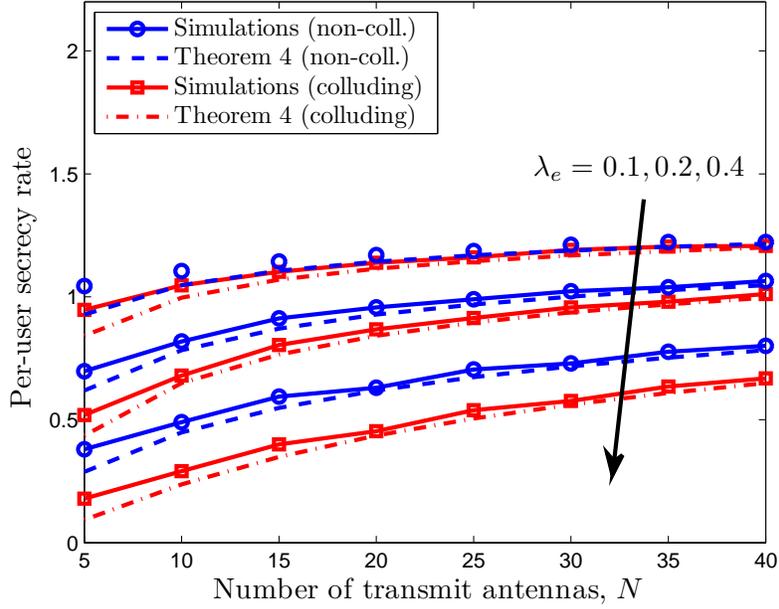}
\caption{Comparison between the simulated ergodic per-user secrecy rate $\mathbb{E}[R_k]$ under non-colluding and colluding eavesdroppers, and the large-system results $\Ro$ from Theorem 4, for a network load $\beta=1$, an SNR $\rho=10$dB, and various values of $\lambda_e$.}
\label{fig:SG_Rs_vs_M}
\end{figure}

In Fig. \ref{fig:SG_Rs_vs_M} we compare the simulated ergodic per-user secrecy rate under non-colluding and colluding eavesdroppers, to the large-system results from Theorem 4, for $\beta=1$, $\rho=10$dB, $\xi=\xiBCCo$, and various values of $\lambda_e$. We note that the accuracy of the large-system analysis increases with $N$. Moreover, we observe that the expectation of the per-user secrecy rate increases with $N$, and this benefit is more for larger values of $\lambda_e$. This happens because the mean received power at each external eavesdropper scales as $\frac{1}{\beta N}$, hence having more transmit antennas makes the system more robust against external eavesdroppers.

\begin{figure}
\centering
\includegraphics[width=\figwidth]{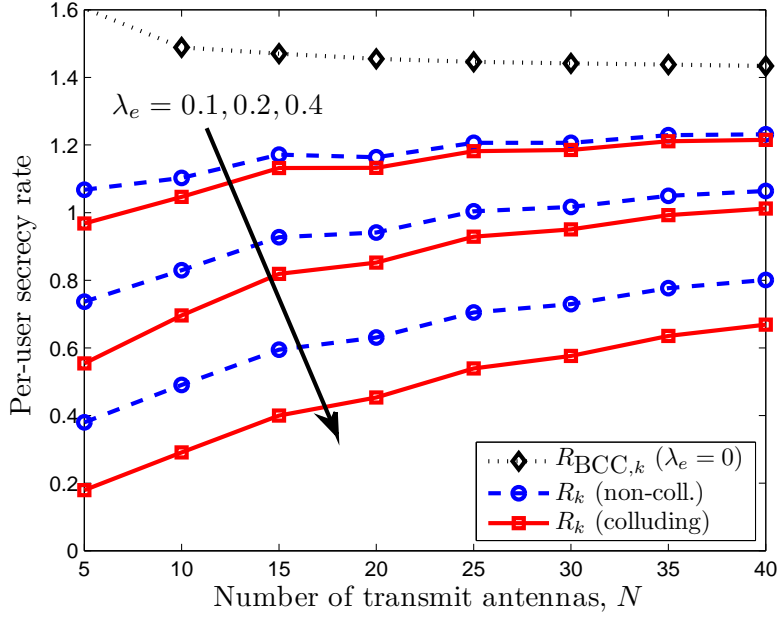}
\caption{Comparison between the simulated ergodic per-user secrecy rates of: (i) the BCCE with non-colluding eavesdroppers, (ii) the BCCE with colluding eavesdroppers, and (iii) the BCC without external eavesdroppers, for a network load $\beta=1$, an SNR $\rho=10$dB, and various values of $\lambda_e$.}
\label{fig:BCC_BCCE}
\end{figure}

In Fig. \ref{fig:BCC_BCCE} we compare the simulated per-user secrecy rate of (i) the BCCE with non-colluding eavesdroppers, (ii) the BCCE with colluding eavesdroppers, and (iii) the BCC without external eavesdroppers, for $\beta=1$, $\rho=10$dB, $\xi=\xiBCCo$, and various values of $\lambda_e$. We note that in the BCC, the per-user secrecy rate is almost constant with $N$, for a fixed network load $\beta$. On the other hand, the per-user secrecy rate of the BCCE increases with $N$. Again, this happens because the mean received power at each external eavesdropper scales as $\frac{1}{\beta N}$, hence having more transmit antennas makes the system more robust against external eavesdroppers. We also note that for higher densities of eavesdroppers $\lambda_e$, larger values of $N$ are required to achieve a given per-user secrecy rate of the BCCE. More precisely, increasing $\lambda_e$ by a factor $2$, requires increasing $N$ by a factor 4. Moreover, the collusion of external eavesdroppers does not affect the scaling law of the mean rate. These observations are consistent with Remark 2.

\begin{figure}
\centering
\includegraphics[width=\figwidth]{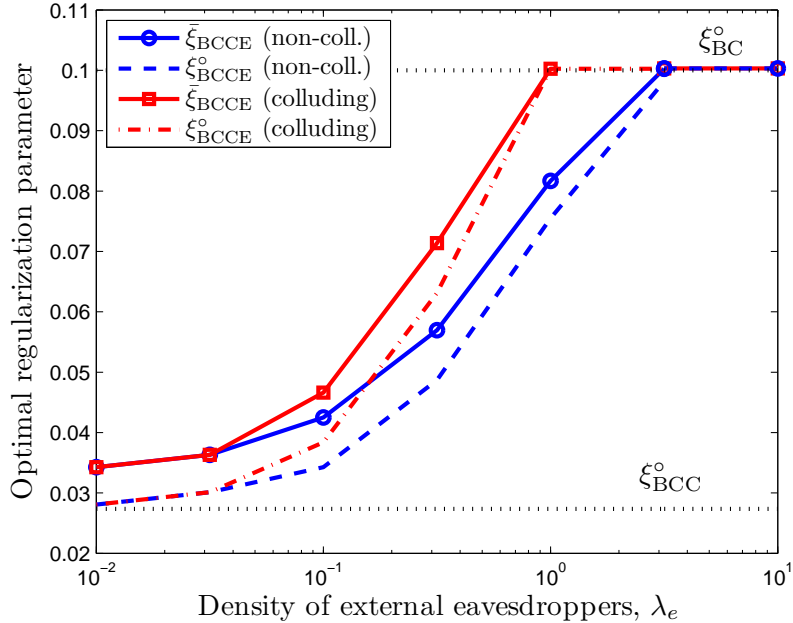}
\caption{Comparison between the large-system regularization parameter $\xiBCCEo$ in (\ref{eqn:xi_BCCE}) and the value $\xiBCCEb$ that maximizes the average simulated secrecy sum-rate $S$ for a finite system with $N=10$ transmit antennas, a network load $\beta=1$, and an SNR $\rho=10$dB.}
\label{fig:SG_xi_vs_lambdae}
\end{figure}

Fig. \ref{fig:SG_xi_vs_lambdae} compares the large-system regularization parameter $\xiBCCEo$ given by (\ref{eqn:xi_BCCE}) to the value $\xiBCCEb$ that maximizes the average simulated secrecy sum-rate $S$, for a finite system with $N=10$, $\beta=1$, and $\rho=10$dB. The figure shows that for low densities of eavesdroppers $\lambda_e$, $\xiBCCEo$ tends to $\xiBCCo=0.0273$, whereas for high densities $\lambda_e$, it tends to $\xiBCo=0.1$. These observations are consistent with Lemma 5. The finite-system parameter $\xiBCCEb$ follows a similar trend. We note that both $\xiBCCEo$ and $\xiBCCEb$ are smaller in the case of non-colluding eavesdroppers, and this can be explained as follows. A smaller value of $\xi$ generates a smaller information leakage to the malicious users. Therefore, it is especially desirable to have a smaller $\xi$ when the malicious users are the main concern, i.e., when their SINR is larger than the SINR at the external eavesdroppers, and this is more likely to happen when the external eavesdroppers are not colluding.

\begin{figure}
\centering
\includegraphics[width=\figwidth]{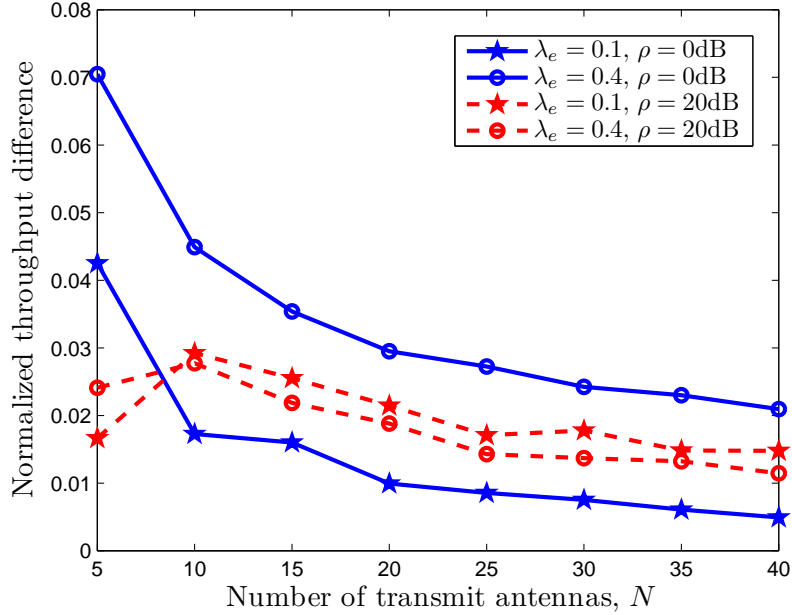}
\caption{Normalized mean secrecy sum-rate difference between using $\xiBCCEs$, that maximizes each realization of the secrecy sum-rate $S$, and $\xiBCCEo$, obtained from large-system analysis in (\ref{eqn:xi_BCCE}), under colluding eavesdroppers, for a network load $\beta=1$, various values of the density of eavesdroppers $\lambda_e$, and various values of the SNR $\rho$.}
\label{fig:Rs_loss_vs_M}
\end{figure}

Fig. \ref{fig:Rs_loss_vs_M} shows that using the regularization parameter $\xiBCCEo$, obtained from large-system analysis, does not cause a significant loss compared to using the optimal parameter $\xiBCCEs$, optimized for
each realization of the channels and of the locations of the external eavesdroppers. The figure shows the mean secrecy sum-rate difference $S(\xiBCCEs)-S(\xiBCCEo)$ normalized by the mean optimal $S(\xiBCCEs)$, simulated for finite-size systems, $\beta=1$, various values of the density of eavesdroppers $\lambda_e$, and various values of the SNR $\rho$. Fig. \ref{fig:Rs_loss_vs_M} was obtained for colluding eavesdroppers, but similar results were obtained for non-colluding eavesdroppers. We note that calculating the optimal value $\xiBCCEs$ requires the base station to know (i) the channels $\mathbf{H}$ of all users, (ii) the realization of the PPP $\Phi_e$, i.e., the locations of all external eavesdroppers, and (iii) the channels $\mathbf{h}^{\dagger}_e$ of all external eavesdroppers. On the other hand, calculating $\xiBCCEo$ does not require the knowledge of any of these quantities. We observe that the normalized mean secrecy sum-rate difference is less than $7\%$ for all values of $N$, $\lambda_e$, and $\rho$, and it decreases when $N$ grows, e.g., falling under $3\%$ for $N=20$. As a result, one can avoid the calculation of $\xiBCCEs$ for every realization of $\mathbf{H}$, $\Phi_e$, and $\mathbf{h}^{\dagger}_e$, and $\xiBCCEo$ can be used with only a small loss of performance.

\section{Conclusion}
In this paper, we considered the broadcast channel with confidential messages and external eavesdroppers (BCCE), where a multi-antenna base station simultaneously communicates to multiple malicious users, in the presence of randomly located external eavesdroppers. We showed that, irrespective of the collusion strategy at the external eavesdroppers, a large number of transmit antennas $N$ drives both the probability of secrecy outage and the rate loss due to the presence of external eavesdroppers to zero. Increasing the density of eavesdroppers $\lambda_e$ by a factor $n$, requires $n^2$ as many antennas to meet a given probability of secrecy outage and a given mean secrecy rate. Using the developed analysis, we clearly established the importance of the number of transmit antennas at the BS to make communications robust against malicious users and external eavesdropping nodes. Investigating the secrecy rates in a cellular scenario, where multiple base stations generate inter-cell interference and malicious users of neighboring cells can cooperate, could be an interesting future research direction.
\appendices

\section{}

\begin{IEEEproof}[Proof of Lemma \ref{Lemma1}]
If $\gLk \leq \gMk$, then $R_k$ in (\ref{eqn:Rs_e}) is zero w.p. 1. If $\gLk > \gMk$, we have for non-colluding eavesdroppers $\gEk = \underset{e}{\max} \!\enspace \gek$, therefore
\begin{align}
\Ok &= \mathbb{P} \left( \gEk \geq \gLk \: \Big| \: \gLk \right) = 1 - \EPhi \left[ \prod_{x \in \Phi_e} {\mathbb{P} \Big( \gxk < \gLk \: \Big| \: \gLk\Big) } \right] \nonumber\\
&= 1 - \EPhi \left[ \prod_{x \in \Phi_e} {\left[1 - P \Big( \gxk \geq \gLk \: \Big| \: \gLk \Big) \right]} \right] \stackrel{(a)}{=} 1 - \EPhi \left[ \prod_{x \in \Phi_e} {\left[ 1 - \mathrm{exp}\Big( - N \beta \sigma^2 \gLk \|x\|^{\eta} \Big) \right]} \right] \nonumber\\
& \stackrel{(b)}{=} 1 - \mathrm{exp}\left[ -2 \pi \lambda_e \int_0^{\infty} { y \, \mathrm{exp}\Big( - N \beta \sigma^2 \gLk \, y^{\eta} \Big) dy } \right] \nonumber\\
& \stackrel{(c)}{=} 1 - \mathrm{exp}\left[ -\pi \lambda_e \int_0^{\infty} { \mathrm{exp}\Big( - N \beta \sigma^2 \gLk \, u^{\frac{\eta}{2}} \Big) du } \right] \stackrel{(d)}{=} 1 - \mathrm{exp}\left[ - \frac{2 \pi \lambda_e}{\eta (N \beta \sigma^2 \gLk)^{\frac{2}{\eta}}} \int_0^{\infty} { e^{-t} t^{\frac{2}{\eta}-1} dt } \right] \nonumber\\
& \stackrel{(e)}{=} 1 - \mathrm{exp}\left[ - \frac{2 \pi \lambda_e \Gamma\left(\frac{2}{\eta}\right)}{\eta (N \beta \sigma^2 \gLk)^{\frac{2}{\eta}}} \right]
\label{eqn:Pout_2}
\end{align}
where (a) follows from the distribution of $\gek$, (b) follows by using $\|x\|=y$, by applying the probability generating functional (PGFL) for the PPP $\Phi_e$, given by \cite{Stoyan96}
\begin{equation}
\EPhi \left[ \prod_{x \in \Phi_e} {f(x) } \right] = \mathrm{exp}\left\{ -\int_{\mathbb{R}^2} {\left[1-f(x)\right] \lambda_e dx } \right\}
\end{equation}
and by changing to polar coordinates. Moreover, in (c) we have used $u = y^2$, in (d) we have used $t = M \beta \sigma^2 \gLk u^{\frac{\eta}{2}}$, and (e) follows from the definition of the gamma function
\begin{equation}
\Gamma(z) \triangleq \int_0^{\infty} {t^{z-1} e^{-t} dt}.
\end{equation}
\end{IEEEproof}

\section{}

\begin{IEEEproof}[Proof of Lemma \ref{Lemma2}]
If $\gLk \leq \gMk$, then $R_k$ in (\ref{eqn:Rs_e}) is zero with probability one. If $\gLk > \gMk$, we have for the eavesdropper nearest to the BS
\begin{align}
\Ok &= \mathbb{P} \left( \gEk \geq \gLk \: \Big| \: \gLk \right) \nonumber\\
&= \int_0^{\infty} \mathbb{P} \Big( \gEk \geq \gLk \: \Big| \: \gLk, \|E\|=x \Big) f_{\|E\|}(x) dx \nonumber\\
&= \int_0^{\infty} \mathbb{P} \Big( \frac{x^{-\eta}}{\sigma^2} \left| \mathbf{h}_E^{\dagger} \mathbf{w}_k \right|^2 \geq \gLk \: \Big| \: \gLk, \|E\|=x \Big) f_{\|E\|}(x) dx \nonumber\\
&= \int_0^{\infty} \mathbb{P} \Big( \left| \mathbf{h}_E^{\dagger} \mathbf{w}_k \right|^2 \geq \sigma^2 \gLk x^{\eta} \: \Big| \: \gLk, \|E\|=x \Big) f_{\|E\|}(x) dx \nonumber\\
& \stackrel{(a)}{=} \int_0^{\infty} {\mathrm{exp}\Big( - N \beta \sigma^2 \gLk x^{\eta} \Big) f_{\|E\|}(x) dx} \nonumber\\
& \stackrel{(b)}{=} 2 \pi \lambda_e \int_0^{\infty} { x \, \mathrm{exp}\Big( - N \beta \sigma^2 \gLk \, x^{\eta} - \lambda_e \pi x^2 \Big) dx},
\label{eqn:P1_appendix}
\end{align}
where (a) holds because $\left| \mathbf{h}_E^{\dagger} \mathbf{w}_k \right|^2 \sim \textrm{exp}(\frac{1}{N\beta})$, and (b) holds because the distance $\|E\|$ between the base station and the nearest eavesdropper $E$ has distribution \cite{Haenggi05}
\begin{equation}
f_{\|E\|}(x) = 2 \lambda_e \pi x \exp(- \lambda_e \pi x^2).
\end{equation}
For a path loss exponent $\eta=4$, (\ref{eqn:P1_appendix}) reduces to
\begin{align}
\Ok &= 2 \pi \lambda_e \int_0^{\infty} { x \, \mathrm{exp}\Big( - N \beta \sigma^2 \gLk \, x^4 - \lambda_e \pi x^2 \Big) dx} \nonumber\\
& \stackrel{(c)}{=} \pi \lambda_e \int_0^{\infty} { \mathrm{exp}\Big( - N \beta \sigma^2 \gLk \, u^2 - \lambda_e \pi u \Big) du} \nonumber\\
& \stackrel{(d)}{=} \frac{\pi^{\frac{3}{2}} \lambda_e}{2 \sqrt{ N \beta \sigma^2 \gLk} } \mathrm{exp} \left[ \frac{( \pi \lambda_e)^2}{4 N \beta \sigma^2 \gLk } \right] \textrm{erfc} \left( \frac{\pi \lambda_e}{2 \sqrt{N \beta \sigma^2 \gLk}} \right)
\end{align}
where in (c) we have used $u = x^2$, and (d) follows from
\begin{equation}
\int_0^{\infty} \textrm{exp} (-ax^2-bx) dx = \frac{1}{2} \sqrt{\frac{\pi}{a}} \, \textrm{exp}\left(\frac{b^2}{4a}\right)  \, \textrm{erfc} \left(\frac{b}{2\sqrt{a}}\right).
\end{equation}
\end{IEEEproof}

\section{}

\begin{IEEEproof}[Proof of Lemma \ref{Lemma3}]
For the case of colluding eavesdroppers, the Laplace transform of the SINR is \cite{HaenggiInvited}
\begin{align}
\mathcal{L}_{\gEk}(s) &= \mathbb{E} \left[ \mathrm{exp} \left( -\frac{s}{\sigma^2} \sum_{x \in \Phi_e} \|x\|^{-\eta} \left| \mathbf{h}_x^{\dagger} \mathbf{w}_k \right|^2 \right) \right] \nonumber\\
& \stackrel{(a)}{=} \textrm{exp} \left\{ -2 \pi \lambda_e \int_{\mathbb{R}^2} \mathbb{E}_{\mathbf{h}} \left[ 1 - \textrm{exp} \left( -\frac{s}{\sigma^2} \left| \mathbf{h}_x^{\dagger} \mathbf{w}_k \right|^2 \|x\|^{-\eta} \right) \right] x \, dx \right\} \nonumber\\ 
& \stackrel{(b)}{=} \textrm{exp} \left\{ - \pi \lambda_e \, \mathbb{E}_{\mathbf{h}} \left[ \left| \frac{1}{\sigma}\mathbf{h}_x^{\dagger} \mathbf{w}_k \right|^{\frac{4}{\eta}} \right] \Gamma\left(1-\frac{2}{\eta}\right) s^{\frac{2}{\eta}} \right\} \nonumber\\ 
& \stackrel{(c)}{=} \textrm{exp} \left\{ - \pi \lambda_e \, \left(N \beta \sigma^2\right)^{-\frac{2}{\eta}} \, \Gamma\left(1+\frac{2}{\eta}\right) \Gamma\left(1-\frac{2}{\eta}\right) s^{\frac{2}{\eta}} \right\}
\label{eqn:Laplace}
\end{align}
where (a) holds since $\Phi_e$ is a PPP \cite{HaenggiInvited}, (b) follows since the fading is independent of the point process, and (c) follows since $\left| \mathbf{h}_x^{\dagger} \mathbf{w}_k \right|^2 \sim \textrm{exp}(\frac{1}{N\beta})$.
Under a path loss exponent $\eta=4$, (\ref{eqn:Laplace}) reduces to
\begin{equation}
\mathcal{L}_{\gEk}(s) = \textrm{exp} \left( - \frac{\pi^2 \lambda_e}{2} \sqrt{\frac{s}{N \beta \sigma^2}} \right).
\end{equation}
By inverse transform one can obtain the distribution function \cite{Sousa90}
\begin{equation}
f_{\gEk}(y) = \frac{\pi^{\frac{3}{2}}\lambda_e y^{-\frac{3}{2}}}{4 \sqrt{N \beta \sigma^2}} \textrm{exp} \left( -\frac{\pi^4 \lambda_e^2}{16 N \beta \sigma^2 y} \right),
\label{eqn:f_coll}
\end{equation}
which integrated yields the cumulative distribution function
\begin{equation}
F_{\gEk}(y) = \textrm{erfc} \left[ \frac{\pi^2 \lambda_e}{4 \sqrt{N\beta\sigma^2 y}} \right],
\end{equation}
from which the secrecy outage probability in (\ref{eqn:Pc}) can be calculated as $\Ok = F_{\gEk}(\gLk)$.
\end{IEEEproof}

\section{}

\begin{IEEEproof}[Proof of Lemma \ref{Lemma4}]
We note from (\ref{eqn:Rs_e}) that when $\gLk \leq \gMk$, the secrecy rate $R_k$ is zero $\forall \, \gEk$. When $\gLk > \gMk$, the mean secrecy rate is given by
\begin{align}
\EPhi \left[ R_k | \gLk \!>\! \gMk \right] &= \EPhi \left[ \max \left[ \log_2 \Big( 1 + \gLk \Big) - 
\log_2 \Big( 1 + \max \left(  \gEk , \gMk \right)
\Big), 0 \right] \right] \nonumber\\
&= \EPhi \left[ \left[ \log_2 \Big( 1 + \gLk \Big) - \log_2 \Big( 1 + \max \left(  \gEk , \gMk \right) \Big) \right] \indic_{(\gamma_{E,k} < \gLk)} \right] \nonumber\\
&= \EPhi \left[ \log_2 \Big( 1 + \gLk \Big) \indic_{(\gamma_{E,k} < \gLk)} - \log_2 \Big( 1 + \max \left(  \gEk , \gMk \right) \Big) \indic_{(\gamma_{E,k} < \gLk)} \right] \nonumber\\
&= \mathbb{P} \left( \gamma_{E,k} < \gLk \right) \log_2 \Big( 1 + \gLk \Big) \!-\! \EPhi \left[  \log_2 \Big( 1 \!+\! \max \left(  \gEk , \gMk \right) \Big) \indic_{(\gamma_{E,k} < \gLk)} \right] \nonumber\\
&= \mathbb{P} \left( \gamma_{E,k} < \gLk \right) \log_2 \Big( 1 + \gLk \Big) \nonumber\\
&\quad - \EPhi \left[ \log_2 \Big( 1 + \gMk \Big) \indic_{(\gamma_{E,k} < \gMk)} + \log_2 \Big( 1 + \gEk \Big) \indic_{(\gMk < \gamma_{E,k} < \gLk)} \right] \nonumber\\
&= \mathbb{P} \left( \gamma_{E,k} < \gLk \right) \log_2 \Big( 1 + \gLk \Big) - \mathbb{P} \left( \gamma_{E,k} < \gMk \right) \log_2 \Big( 1 + \gMk \Big) \nonumber\\
&\quad - \int_{\gMk}^{\gLk} {\log_2(1+y) f_{\gEk}(y) \, dy} \nonumber\\
&= \log_2 \Big( 1 + \gLk \Big)^{1-\Ok} - \log_2 \Big( 1 + \gMk \Big)^{1-\Pk} - \int_{\gMk}^{\gLk} {\log_2(1+y) f_{\gEk}(y) \, dy} \nonumber\\
&= \log_2 \frac{\Big( 1 + \gLk \Big)^{1-\Ok}}{\Big( 1 + \gMk \Big)^{1-\Pk}} - \int_{\gMk}^{\gLk} {\log_2(1+y) f_{\gEk}(y) \, dy}
\end{align}
where (i) $\indic_{(\cdot)}$ is the indicator function, (ii) $\Ok \triangleq \mathbb{P} \left( \gamma_{E,k} \geq \gLk \right)$ is given by the secrecy outage probability; (iii) $\Pk \triangleq \mathbb{P} \left( \gamma_{E,k} \geq \gMk \right)$ is the probability that the SINR at the external eavesdroppers is greater than or equal to the SINR at the malicious users, given in (\ref{eqn:Pk}) and obtained by calculations similar to the ones in Lemma 1 and Lemma 2; and (iv) $f_{\gEk}(y)$ is the distribution of the SINR at the external eavesdroppers, given by (\ref{eqn:f_coll}) for colluding eavesdroppers, and by
\begin{equation}
f_{\gEk}(y) = \frac{\partial \mathbb{P}\left( \gEk < y\right)}{\partial y}  = \frac{\pi^{\frac{3}{2}}\lambda_e y^{-\frac{3}{2}}}{4 \sqrt{N \beta \sigma^2}} \textrm{exp} \left( -\frac{\pi^{\frac{3}{2}} \lambda_e}{2 \sqrt{N \beta \sigma^2 y}} \right)
\end{equation}
for non-colluding eavesdroppers.
\end{IEEEproof}

\section{}

\begin{IEEEproof}[Proof of Corollary 3]
For $\gLo \leq \gMo$, we have $\RBCCo=0$ and $\Ro=0$, therefore $\Delta_e=0$. For $\gLo > \gMo$ and fixed $\xi$, irrespective of the cooperation strategy at the eavesdroppers, we have
\begin{align}
\Delta_e &= \Oo \log(1+\gLo) - \Po \log(1+\gMo) + \int_{\gMo}^{\gLo} {\log_2(1+y) f_{\gEk}(y) \, dy} \nonumber\\
&\stackrel{(a)}{\leq} \Oo \RBCCo + \frac{\mu \lambda_e}{2 \sqrt{N}} \int_{\gMo}^{\gLo} {y^{-\frac{1}{2}} \, dy} \nonumber\\
&= \left[1-\textrm{exp}\left(-\frac{\mu \lambda_e}{\sqrt{N \gLo}}\right) \right] \RBCCo + \frac{\mu \lambda_e}{\sqrt{N}} \left( \sqrt{\gLo} - \sqrt{\gMo} \right) \nonumber\\
&\leq \frac{\mu \lambda_e}{\sqrt{N \gLo}} \RBCCo + \frac{\mu \lambda_e}{\sqrt{N}} \left( \sqrt{\gLo} - \sqrt{\gMo} \right) = \mu \left[ \frac{\RBCCo}{\sqrt{\gLo}} + \left( \sqrt{\gLo} - \sqrt{\gMo} \right) \right] \frac{\lambda_e}{\sqrt{N}}
\end{align}
where (a) holds because $\Po > \Oo$, $\log_2(1+y) \leq y$, and $f_{\gEk}(y) \leq \frac{\mu \lambda_e y^{-\frac{3}{2}}}{2 \sqrt{N}}$.
\end{IEEEproof}
\ifCLASSOPTIONcaptionsoff
  \newpage
\fi
\bibliographystyle{IEEEtran}
\bibliography{IEEEabrv,Bib_Giovanni}
\end{document}